\documentclass[journal,10pt]{IEEEtran}
\usepackage{cite,amsmath,amssymb,amsfonts,algorithmic,multirow,graphicx,textcomp,xcolor,booktabs,listings,url,threeparttable,subfigure}
\usepackage{epstopdf}
\usepackage{makecell}
\usepackage[ruled,linesnumbered]{algorithm2e}

\begin{document}
\title{EdgeShard: Efficient LLM Inference via Collaborative Edge Computing
}
\author{Mingjin~Zhang,
        ~Jiannong~Cao,~\IEEEmembership{Fellow,~IEEE,}
        ~Xiaoming~Shen,
        ~Zeyang~Cui
\IEEEcompsocitemizethanks{\IEEEcompsocthanksitem M. Zhang, J. Cao, and X. Shen, and Z. Cui are with the Department of Computing, The Hong Kong Polytechnic University, Hong Kong. 
\protect\\ E-mail: \{csmzhang, csjcao\}@comp.polyu.edu.hk
}
}

\markboth{Journal of \LaTeX\ Class Files,~Vol.~14, No.~8, August~2021}%
{Shell \MakeLowercase{\textit{et al.}}: A Sample Article Using IEEEtran.cls for IEEE Journals}

\maketitle

\begin{abstract}
Large language models (LLMs) have shown great potential in natural language processing and content generation. However, current LLMs heavily rely on cloud computing, leading to prolonged latency, high bandwidth cost, and privacy concerns. Edge computing is promising to address such concerns by deploying LLMs on edge devices, closer to data sources. Some works try to leverage model quantization to reduce the model size to fit the resource-constraint edge devices, but they lead to accuracy loss. Other works use cloud-edge collaboration, suffering from unstable network connections. In this work, we leverage collaborative edge computing to facilitate the collaboration among edge devices and cloud servers for jointly performing efficient LLM inference. We propose a general framework to partition the LLM model into shards and deploy on distributed devices. To achieve efficient LLM inference, we formulate an adaptive joint device selection and model partition problem and design an efficient dynamic programming algorithm to optimize the inference latency and throughput, respectively. Experiments of Llama2 serial models on a heterogeneous physical prototype demonstrate that EdgeShard achieves up to 50\% latency reduction and 2x throughput improvement over baseline methods.
\end{abstract}


\section{Introduction}\label{sec:introduction}
\IEEEPARstart{R}ecently, the emergence of Large Language Models (LLMs) has attracted widespread attention from the public, industry, and academia, representing a significant breakthrough in artificial intelligence (AI). Many players are coming into this field with their advanced models, such as OpenAI's GPT-4 \cite{achiam2023gpt}, Meta's Llama \cite{touvron2023Llama}, and Google's PALM \cite{anil2023palm}. Built on the foundation of transformer architecture \cite{vaswani2017attention}, LLMs are characterized by their massive scale in terms of the number of parameters and the amount of data they are trained on. The scale of LLMs, often numbering in hundreds of billions of parameters, enables the models to capture complex patterns in language and context, making them highly effective at generating coherent and contextually appropriate responses. Such a phenomenon is also known as "intelligence emergence". The outstanding capability of LLMs makes them valuable and well-performed in a wide range of applications, from ChatBot and content generation (e.g., text summation and code generation) to assisting tools of education and research.  

\begin{figure}[t]
	\centering
	\includegraphics[width=0.9\linewidth]{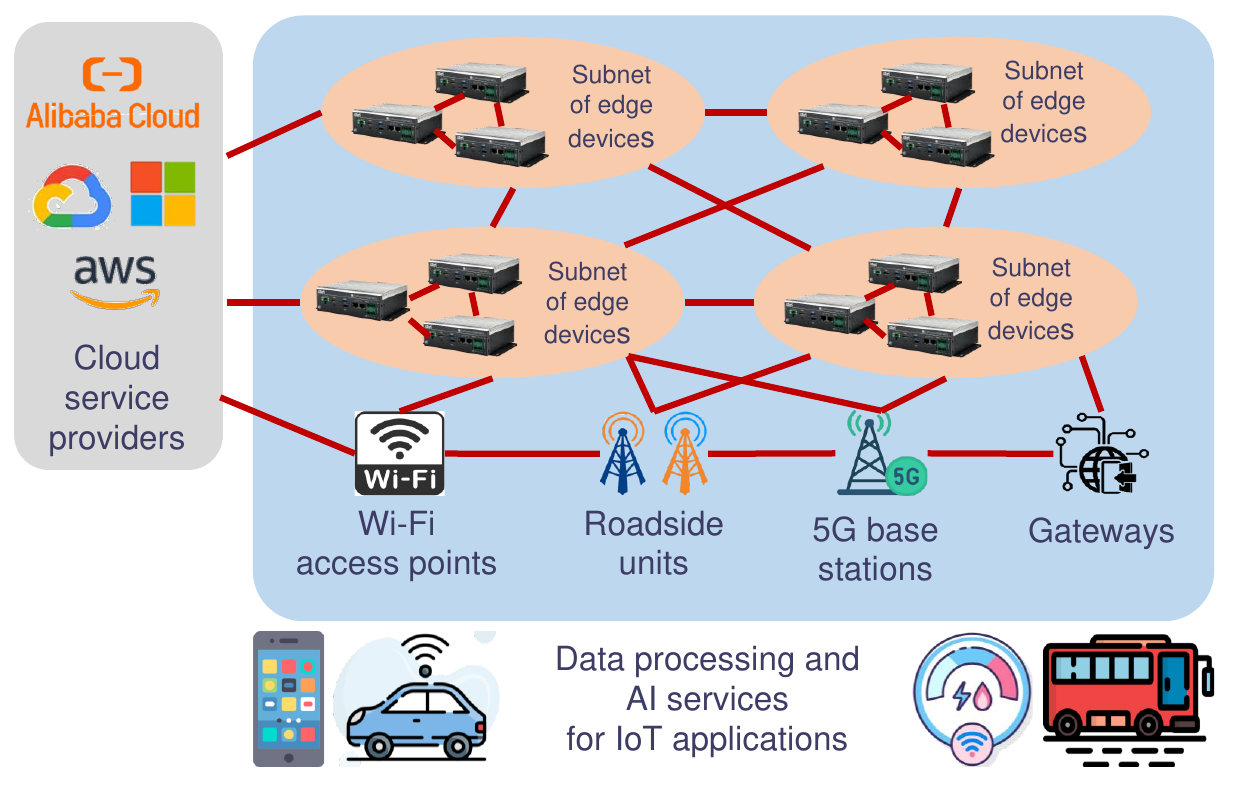}
	\caption{Collaborative edge computing integrates the computing resources of ubiquitous geo-distributed devices for jointly performing computational tasks, with great benefits of enlarged resource pool, low-latency data processing, flexible device access, and expanded service region.}
	\label{f:cec_framework}
\end{figure}

However, current LLMs heavily rely on cloud computing, suffering from long response time, high bandwidth cost, and privacy concerns \cite{shi2016edge}. Firstly, the reliance on cloud computing hampers the capability for rapid model inference necessary for real-time applications such as robotics control, navigation, or exploration, where immediate responses are crucial. Secondly, the transmission of large amounts of data, including texts, video, images, audio, and IoT sensing data, to the cloud data centers leads to substantial bandwidth consumption and immense strain on the network architecture. Thirdly, cloud-based LLMs raise significant privacy issues, especially when handling sensitive data of hospitals and banks, as well as personal data like text inputs and photos on mobile phones.

Edge computing is a promising solution to address the aforementioned challenges by deploying LLMs on edge devices (e.g., edge servers, edge gateways, and mobile phones) at the network edge closer to the data sources \cite{chen2019deep}. However, LLMs are computation-intensive and resource-greedy. For example, the inference of a full-precision Llama2-7B model requires at least 28GB memory, which may exceed the capacity of most edge devices. Some works leverage model quantization \cite{shen2023agile,frantar2022gptq,frantar2022optq,lin2023awq,xiao2023smoothquant,shen2024edgeqat} to reduce the model size to fit into the resource-constraint edge devices. However, they often lead to accuracy loss. Other works tend to use cloud-edge collaboration \cite{wang2023privatelora,chen2023netgpt}, which partitions the LLMs into two sub-models and offloads part of the computation workload to the powerful cloud servers with high-end GPUs. However, the latency between edge devices and cloud servers is usually high and unstable. 

Alternatively, we have witnessed the continuous growth of the computing power of edge in recent years, and a large number of edge servers and edge clouds have been deployed at the network edge, leaving significant resources to be used. Collaborative edge computing (CEC) \cite{zhang2022eaas,zhang2022ents} is hence proposed recently to integrate the computing resources of geo-distributed edge devices and cloud servers for efficient resource utilization and performance optimization. As shown in Fig.~\ref{f:cec_framework}, ubiquitous and distributed edge devices and cloud servers are connected and form a shared resource pool, collaboratively providing instant data processing and AI services. CEC is different from existing edge computing research. Existing edge computing research focuses on the vertical collaboration among cloud, edge, and end devices, while neglecting horizontal edge-to-edge collaborations, suffering from unoptimized resource utilization, restricted service coverage, and uneven performance.

Motivated by the vision of CEC, we propose a general LLM inference framework, named EdgeShard, to support efficient collaborative LLM inference on distributed edge devices and cloud servers. For simplicity, we use computing devices below to refer to edge devices and cloud servers. Given a network with heterogeneous computing devices, EdgeShard partitions the LLM into multiple shards and allocates them to judicious devices based on the heterogeneous computation and networking resources, as well as the memory budget of devices. To optimize performance, we formulate a joint device selection and model partition problem and design an efficient dynamic programming algorithm to minimize the inference latency and maximize the inference throughput, respectively. Extensive experiments on a practical testbed show that EdgeShard reduces up to 50\% latency and achieves 2x throughput over on-device and vertical cloud-edge collaborative inference methods.  

Our work is different from those works that partition the LLMs and allocate to multiple GPUs in cloud data centers, such as Gpipe \cite{huang2019gpipe} and PipeDream \cite{narayanan2019pipedream}. Deploying LLM at edge computing is vastly different from that in the cloud. First, cloud servers are usually with homogeneous GPUs, while edge devices are with heterogeneous computation capabilities in nature. Second, modern cloud GPUs for LLMs are usually connected by high-bandwidth networks, such as InfiniBand and Nvlinks, while edge devices are connected with heterogeneous and low-bandwidth networks. For example, the bandwidth of NVlinks can go up to 600GB/s, while the bandwidth among edge devices ranges from dozens of Kbps to 1000Mbps. The solution of LLMs deployment designed for cloud data centers neglect the heterogeneous and resource-constrained edge computing environment.  

Our contributions are three folds.
\begin{itemize}
	\item First, we propose a general LLM inference framework for deploying LLMs in the edge computing environment, which enables the collaborative inference among heterogeneous edge devices and cloud servers.
	\item Further, we quantitatively study how to select computing devices and how to partition the LLM for optimized performance. We mathematically formulate a joint device selection and model partition problem, and propose a dynamic programming algorithm to optimize the latency and throughput, respectively.
	\item We also evaluate the performance of EdgeShard with state-of-the-art Llama2 serial models on a physical testbed. Experimental results show EdgeShard remarkably outperforms various baseline methods.
\end{itemize}

\section{Preliminaries and Motivations}

\begin{figure}[t]
	\centering
	\includegraphics[width=0.8\linewidth]{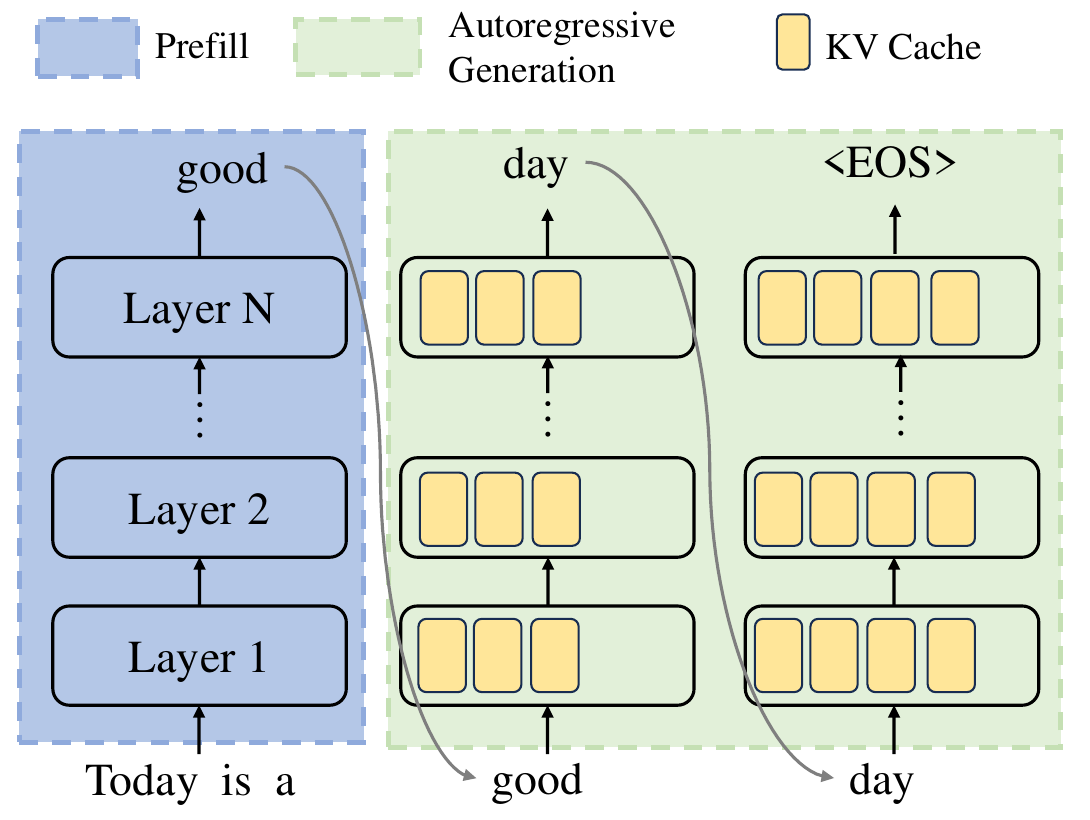}
	\caption{LLM inference has an autoregressive nature.}
	\label{f:llm_inference}
\end{figure}

\begin{table}[t]
	\centering
	\caption{Minimum Memory usage of LLMs inference and memory capacity of Edge Devices.}
	\label{Tab: Device and llm mem}
	\begin{tabular}{lcccc}
		\toprule
		Model & \makecell[c]{Full\\Precision}  & 8-bit & 4-bit & \makecell{Edge\\Devices} \\
		\midrule
		Llama2-7B  & 28GB & 7GB & 3.5GB & Smartphone(6-12GB)\\
		Llama2-13B & 52GB & 13GB & 6.5GB & Jetson Orin(8-16GB)\\
		Llama2-70B & 280GB & 70GB & 35GB &  Jetson AGX(32-64GB)\\
		\bottomrule
	\end{tabular}
\end{table}

\textbf{Generative LLM Inference}. LLMs generally refer to decoder-based transformer models with billions of parameters. Different from encoder-based architecture like BERT \cite{devlin2018bert}, whose inference process is single phase, the process of LLM inference is iterative and typically involves two phases: the prompt processing phase and the autoregressive generation. The prompt processing phase is also known as prefill. 

In the prompt processing phase, the model takes the user initial token $(x_{1},...,x_{n})$ as input and generates the first new token $x_{n+1}$ by computing the probability $P(x_{n+1} \mid x_{1},...,x_{n})$. 

In the autoregressive generation phase, the model generates one token at a time, based on both the initial input and the tokens it has generated so far. This phase generates tokens sequentially for multiple iterations until a stopping criterion is met, i.e., either when generating an end-of-sequence (EOS) token or reaching the maximum number of tokens specified by user or constrained by the LLM.

\begin{figure*}[t]
	\includegraphics[width=0.9\linewidth]{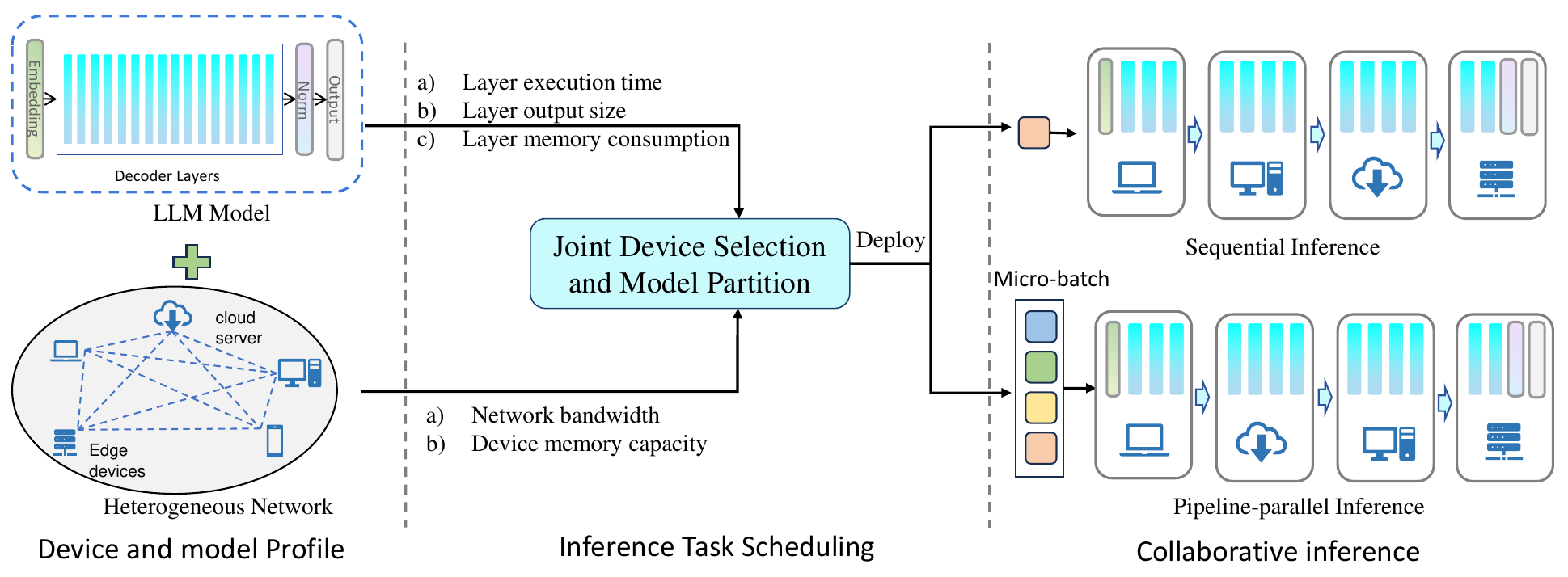} 
	\centering
	\caption{Framework of EdgeLLM. It consists of three stages: offline profiling, task scheduling optimization, and online collaborative LLM inference.}
	\label{framework}
\end{figure*}

As shown in Fig.~\ref{f:llm_inference}, suppose the LLM model has $N$ layers, which will take a sequence of input tokens and run all layers to generate a token in a one-by-one manner. In the prefill phase, the model takes the input ("Today is a") at once, and the first generated token is "good." In the autoregressvie generation phase, the model first takes ("Today is a good") as input and generates the next token ("day"). It then takes ("Today is a good day") as input and generates the next token ("EOS"), which indicates the end of the generation. Since a token generated is determined by all its previous token in a sequence, LLMs utilize Key-Value caching (KV caching) to avoid repetitive computation, storing past computations to expedite responses, thereby reducing computational workload and improving response times. The time to generate a token in the prefill stage is much higher (usually 10x) than that of in the autoregressive stage, as the prefill stage needs to calculate the KV cahche of all input tokens as initialization. 

\textbf{LLMs are memory-consuming.} A single edge device may not have sufficient memory to accommodate a LLM model. Take one of the most popular LLM models, i.e., Llama2, as an example. As shown in Table.~\ref{Tab: Device and llm mem}, Llama2 has three different versions, i.e., 7B, 13B, and 70B. We can see from the Table that the full precision inference of Llama2-7B requires at least $28$GB memory, but the smartphones usually only have 6-12 GB memory, and the Jetson Orin NX has 8-16 GB memory. They are unable to burden the on-device LLM inference. Some works try to use low-precision quantization, e.g., 8 bit and 4 bit. However, it may still exceed the memory capacity of edge devices. For example, the 4-bit inference of Llama2-70B requires at least 35GB memory, which cannot be accommodated on most edge devices. Moreover, low-precision inference leads to performance degradation. 

In this work, we leverage collaborative edge computing, a computing paradigm where geo-distributed edge devices and cloud servers collaborate to perform computational tasks. Based on that idea, we propose EdgeShard, a general LLM inference framework that allows adaptive device selection and LLM partition over distributed computing devices, to address the high memory requirements and leverage heterogeneous resources to optimize LLM inference.

\section{Collaborative Edge Computing for LLMs}

There are three stages of the framework, including profiling, task scheduling optimization, and collaborative inference. The workflow is shown in Fig.~\ref{framework}.

\textbf{Profiling} is an offline step that profiles the necessary run-time traces for the optimization step and only needs to be done once. Those traces include: 1) the execution time of each layer on different devices; 2) the size of activations and memory consumption for each layer of the LLM model; 3) available memory of each device and the bandwidth among devices. For the execution time of each layer, we profile the time to generate a token in the prefill stage and autoregressive stage, respectively, and take the average. For those devices that may not have efficient memory to hold the full model for performing the profiling, we utilize a dynamic model loading technology, where the model layers are consecutively loaded to fit the constrained memory. The profiling information will then be used to support intelligent task scheduling strategies.

\textbf{Scheduling Optimization}. At the task scheduling optimization stage, the scheduler generates a deployment strategy by determining which device to participate in, how to partition the LLM model in a layer wise, and which device should the model shard be allocated to. The strategy thoroughly considers the heterogeneous resources, the memory budget of devices, and the privacy constraint, and later be applied to selected devices for efficient LLM inference. More details is described in Sec.~\ref{sec:optimization}.

\begin{figure}[t]
	\centering
	\subfigure[Sequential inference]{
		\includegraphics[width=0.48\linewidth]{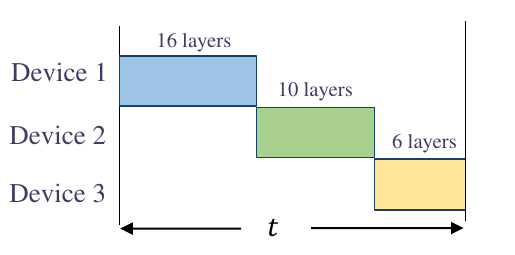}
		%\caption{fig1}
	}
	\hspace{-2em}
	\subfigure[Pipeline parallel inference]{
		\includegraphics[width=0.48\linewidth]{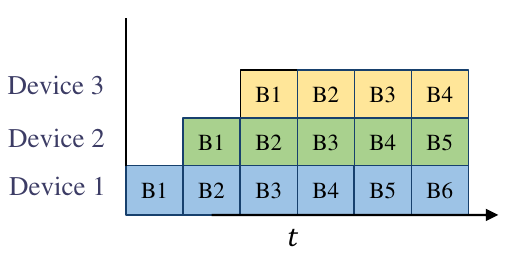}
	}
	\caption{Collaborative LLM inference}
	\label{LLM-inference}
\end{figure}

\textbf{Collaborative inference}. After getting the LLM model partition and allocation strategy, the selected devices will perform the collaborative inference. We pre-allocate memory space for KV cache on each participating device. We consider two cases for the collaborative inference, i.e., sequential inference and pipeline parallel inference. 

In sequential inference, devices take turns to perform the computation with the allocated model shards. As shown in Fig.~\ref{LLM-inference}(a), suppose the LLM model is partitioned into 3 shards and allocated to device 1, 2, and 3, respectively. Device 1 will first process the input data and then send the activations/outputs to device 2, which will process the data and then transmit to device 3. Sequential inference is suitable for serving a single user, such as in smart home scenario, where users' personal devices (e.g., tablet, phones, and smart speaker) collaborate to perform LLM inference. In such scenario, user inputs a prompt and gets the response and then input another prompt. We aims to minimize the latency of sequential inference. 

However, sequential inference is not resource-efficient from the system's perspective. When device 1 is performing computation, device 2 and device 3 are idle. We thus take pipeline parallelism to improve resource utilization. For the pipeline parallel inference as taken in previous work Gpipe \cite{huang2019gpipe} and PipeDream \cite{narayanan2019pipedream} for cloud servers, the input data will first be split into micro-batch and subsequently feed into the system. As depicted in Fig.~\ref{LLM-inference}(b), device 1 first handles data B1 and then transmits intermediate data to device 2. After handling data B1, device 1 immediately goes to handle data B2. In such a pipeline manner, every device is busy with high system resource utilization.

\section{Optimize LLM Inference}\label{sec:optimization}
We consider a general collaborative edge network with heterogeneous devices and bandwidth connection. More specifically, given a set of heterogeneous devices connected with heterogeneous bandwidth, EdgeShard aims to select a subset of devices and partition the LLM into shards, which will be allocated to the selected devices to minimize the inference latency or maximize the throughput.

\textbf{System Model.} LLMs usually have a layered architecture, which consists of an embedding layer, multiple decoder layer, and an output layer. Sizes of parameters and activations (i.e., the output of a layer) vary across layers. We assume the model is with $N$ layers. $O_{i}$ represents the size of activations of layer $i, 0 \leq i \leq N-1$. The memory consumption of a layer $i$ is denoted by $Req_{i}$.

We consider a network consisting of $M$ edge devices and cloud servers. The devices have heterogeneous computation and memory capabilities, and cloud servers are much more powerful than edge devices in terms of computation capability. The memory budget of a device $j$ is $Mem_{j}$. The computing devices are interconnected. Bandwidth between a device $k$ and a device $j$ is $B_{k,j}$, $ 0 \leq k \leq M-1$, $ 0 \leq j \leq M-1$. There is a source node where the input tokens reside. Without loss of generality, we set the source node as node $0$. The main notations used in this paper are shown in Table.~\ref{tab:notation}.

\begin{table}[t]
	\centering
	\caption{List of notations}
	\label{tab:notation}
	\begin{tabular}{|p{1cm}|p{7cm}|}%
		\hline
		\textbf{Symbol} & \textbf{Descriptions} \\%
		\hline
		\hline
		$X_{i,j}$ & binary variable, whether layer $i$ of a model is allocated to device $j$\\%
		\hline
		$t_{comp}^{i,j}$ & computation time of layer $i$ on device $j$ \\%
        \hline
        $t_{comp}^{i \rightarrow m, j}$ & computation time of layer $i$ to layer $m$ on device $j$\\
		\hline
		$t_{comm}^{i-1,k,j}$ & communication time to transmit activations of layer $i-1$ from device $k$ to device $j$ \\%
		\hline
		$DP(i,j)$ & minimal total execution time of the first $i$ layers if layer $i$ is allocated to device $j$ \\%
		\hline
		$g(i,S,k)$ & processing time of the slowest node to process the first $i$ layers with device set $S$ \\%
		\hline
	\end{tabular}
\end{table}

\subsection{Optimize LLM inference latency} 
\textbf{Problem Formulation.} We use a binary variable $X_{i,j}$ to denote the LLM allocation strategy. $X_{i,j}$ equals to $1$ if layer $i$ is allocated to node $j$. Otherwise, $X_{i,j}$ equals to zero. A layer will be and only be allocated to one node. Hence, we have $\sum_{j=0}^{M-1} X_{i,j} = 1, \forall i$. Let $t_{comp}^{i,j}$ denotes the computation time of layer $i$ on node $j$. Suppose layer $i-1$  and layer $i$ are allocated to node $k$ and node $j$, respectively. We use $t_{comm}^{i-1,k,j}$ to denote the communication time to transmit the activations of layer $i-1$ from node $k$ to node $j$. The data transmission time is determined by the output size of a layer and the bandwidth between two nodes. If layer $i-1$ and layer $i$ are on the same node, we assume the transmission time is zero. Hence, we have

\begin{equation}
	t_{comm}^{i-1,k,j} = \begin{cases}
		\frac{O_{i-1}}{B_{k,j}}, \quad & if k \neq j \\
		0, &\text{otherwise.}
	\end{cases}
\end{equation}

The total inference time can thus be calculated by the following equation.

\begin{equation}
	T_{tol} = \sum_{i=0}^{N-1}\sum_{j=0}^{M-1} X_{i,j}*t_{comp}^{i,j} + \sum_{i=1}^{N-1}\sum_{j=0}^{M-1}\sum_{k=0}^{M-1} X_{i-1,k}*X_{i,j}*t_{comm}^{i-1,k,j}
\end{equation}

Hence, the problem of minimizing the LLM inference latency can be formulated as follows, where Eq.~(\ref{latency: privacy}) is the privacy constraint. It shows that the first layer of the LLM model should always be allocated to node $0$, which is set to be the source node with input tokens. In such a case, the raw input data resides on the source node and avoids to be transmitted among computing devices. Eq.~(\ref{latency: memory}) shows that the memory requirements of all the layers allocated to node $j$ cannot exceed its memory budget. 
\begin{equation}
	\min T_{tol}
\end{equation}
\vspace{-1em}
\begin{equation}\label{latency: privacy}
	X_{0,0} = 1 
\end{equation}
\vspace{-1em}
\begin{equation}\label{latency: memory}
	\sum_{i=0}^{N-1} X_{i,j}*Req_{i} \leq Mem_{j} 
\end{equation}

\textbf{Solution.} To minimize the inference latency, we design a dynamic programming algorithm. The intuition is that the minimal execution time of the first $i$ layer is determined by the first $i-1$ layer, which means the optimal solution can be constructed from the optimal results of the sub-problems. It has the optimal sub-problem property, which motivates us to use dynamic programming. 

Let $DP(i,j)$ denote the minimal total execution time of the first $i$ layers after the layer $i$ is allocated to the node $j$. The state transition equation is formulated as:

\begin{equation}\label{infer-transition}
	DP(i,j) = \begin{cases}
        \min\limits_{\substack{k \in M \\ 1 \leq i < N-1}} (DP(i-1,k) + t_{comp}^{i,j} + t_{comm}^{i-1,k,j}) \\
        \min\limits_{\substack{k \in M \\ i = N-1}} (DP(i-1,k) + t_{comp}^{i,j} + t_{comm}^{i-1,k,j} + t_{comm}^{i,j,0}) 
 \end{cases}
\end{equation} 

Where $DP(i-1,k) $ indicates the minimal execution time of the first $i-1$ layers if layer $i-1$ is allocated to device $k$. Eq. (\ref{infer-transition}) shows that $DP(i,j)$ is determined by traversing at all possible nodes of the previous layer and choosing the one that minimizes the execution time of the first $i$ layers. Moreover, due to the autogressive nature of LLM, the generated token needs to be sent back to the source node for next iteration of generation. Hence, for the last layer $N-1$, the communication time not only includes the data transmission time from the $N-2$ layer, but also the transmission time to the source node $t_{comm}^{N-1,j,0}$. Additionally, we initialize $DP(0,0)$ as shown in Eq.~(\ref{latency: initialization}) by considering the privacy constraint.

 \begin{equation}\label{latency: initialization}
	DP(0, 0) = t_{comp}^{0,0}
\end{equation} 

By traversing each layer and each node based on Eq.~(\ref{infer-transition}), we can fill in the dynamic programming table $DP(i,j)$ to track the minimum total execution time to reach each layer. Finally, the minimal total execution time at the last layer can be calculated by Eq.~(\ref{infer-optimal}). We can then get the optimal node allocation for each layer by backtracking $DP(i,j)$.

\begin{equation}\label{infer-optimal}
	min_{j=0,...,M-1}(DP(N-1,j))
\end{equation} 

This method is simple and effective. With dynamic programming, we can quickly traverse the solution space and find the best LLM partition and allocation strategy. The algorithm to find the optimal LLM partition and allocation strategy for minimizing inference latency is shown in Algo.~\ref{algo:latency}.

\begin{algorithm}[t]
	\caption{Joint device selection and LLM partition for optimizing latency}\label{algo:latency}
	\KwIn{A LLM model; Computing device $M$; Profiled traces; bandwidth $B_{k,j}$; }
	\KwOut{the device selection and LLM partition strategy }
	\tcp{initialization}
	Initialize DP table $DP(i,j)=INF$, and choice table $choice(i,j)=NULL$ to record the strategy\;
	Enforce first layer to be allocated to node $0$ by $DP(0, 0) = t_{comp}^{0,0}$ and $choice(0,0)=0$\;
	\tcp{fill in the DP table}
	\For{$i=1$ to $N-1$}{
		\For{$j=0$ to $M-1$}{
			\If{$Mem_{j} \leq Req_{i}$}{
				Continue\;
			}
			\Else{
				\For{$k = 0$ to $M-1$}{
					Calculate the total execution time by Eq.~(\ref{infer-transition}) and assign it to $t_{total}$\;
					\If{$t_{total} \leq DP(i,j) $}{
						Update $DP(i,j)$ by assigning $DP(i,j)=t_{total}$\;
						Update memory $Mem_{j}$\;
						Record allocation plan $choice(i,j) = k$\;
					}
				}
			}
		}
	}
	\tcp{backtrace for allocation strategy}
	Initialize optimal strategy $R$\;
	Find the last selected node $N_{last} = argmin_{j} (DP(N-1,j))$\;
	Add $N_{last}$ to $R$\;
	\For{$i=N-1$ to $0$}{
		Find the previous node $N_{last} = choice(i, N_{last})$\;
		Add $N_{last}$ to $R$\;
	} 
	Reverse $R$\;
	\Return $R$\;
\end{algorithm}

In Algo.~\ref{algo:latency}, we first initialize the dynamic programming table $DP(i,j)$ and choice table $choice(i,j)$ (lines 1-2). We initialize $DP(0,0)$ according to Eq.~(\ref{latency: initialization}). The $choice(i,j)$ records the node $k$ to host the $i-1$ layer. It is the optimal variable of Eq. (\ref{infer-transition}). We then traverse the layers of the large language model from layer $1$. For each layer, we traverse all the computing devices with sufficient memory and calculate the inference time (lines 3-19). After filling the DP table, we can find the minimal $DP(N-1,j)$, which represents the minimal time for executing the LLM model, and the last node to host layer $N-1$. Finally, by backtracing $choice(i,j)$, we get the model partition and allocation strategy $R$ (lines 20-28). The computational complexity of Algo.~\ref{algo:latency} is $O(N \times M \times M)$, where $N$ is the number of layers of the LLM model and $M$ is the number of devices in the network.

\subsection{Optimize LLM inference throughput} 

\textbf{Problem Formulation.} For optimizing throughput, pipeline parallelism is adopted to avoid device idleness. As illustrated before, the computation time of layer $i$ on node $j$ is $t_{comp}^{i,j}$, and if layer $i$ to layer $m$ are all allocated to node $j$, the computation time is indicated by $t_{comp}^{i \rightarrow m, j}$. The data transmission time of the activations of layer $i-1$ from node $k$ to node $j$ is $t_{comm}^{i-1,k,j}$. In pipeline parallel inference, the computation time and communication time can be overlapped to maximize the throughput. Thus, for the inference task, the maximum latency for the device $j$ can be calculated as:
\begin{equation}
	T_{latency}^{j} = \max 
	\left\{
	\begin{array}{lr}
		t_{comp}^{i \rightarrow m, j}\\
		
		t_{comm}^{i-1,k,j} &  
	\end{array}
	\right. \label{T_Period}
\end{equation}

Ideally, for the selected devices, achieving the maximal throughput is equivalent to minimizing the latency of the slowest device. We use $S$ to denote the selected devices, and then the problem of maximizing the inference throughput can thus be formulated as follows, where $j \in S$.

\begin{equation}
	\min \{T_{latency}^{j} | j \in S\}
\end{equation}

\textbf{Solution.} Similar to minimizing the inference latency, the problem of maximizing the throughput also has an optimal sub-problem property. Maximizing the throughput of the first $i$ layer can be deduced from solving the problem of allocating the first $i-1$ layer, which indicates that the optimal solution of the whole problem can be constructed from the sub-problems. We also use dynamic programming to solve the problem.

Let $g(i, S, k)$ denote the minimum time to process the first $i$ layers with the set of used devices $S$, and the device $k$ is the last node to be used, $k \in S$. We use $g(m, S', j)$ to denote the next state to process the first $m$ layers with the set of used devices $S'$, and the device $j$ is the last node to be used, where $0 \leq i < m \leq N-1$, $j \in M \setminus S$, $S'=S\cup \{j\}$.

%% consist of two notations
The state transition equation is formulated in Eq.~ (\ref{transistion_equation}), where $g(m, S', j)$ is determined by the previous state $g(i, S, k), $ and the maximum latency of devive $j$, i.e., the computation time $t_{comm}^{i-1,k,j}$ and the communication time $t_{comp}^{i \rightarrow m, j}$. The final optimal solution $T_{throu}^{opt}$ is the minimum $ g(N-1, S', j)$, where $S' \subseteq M$.

\begin{equation}
	\begin{split}
		\underset{S'=S\cup \{j\}}{g(m, S', j)} = \min_{\substack{0\leq i < m \leq N-1 \\ j \in M \setminus S}}  \max 
		\left\{
		\begin{array}{lr}
			g(i, S, k)\\
			t_{comm}^{i-1,k,j} &  \\
			t_{comp}^{i \rightarrow m, j}  &  
		\end{array}                
		\right. \label{transistion_equation}
	\end{split}
\end{equation}

Additionally, we have constraints when performing state transition. They are the memory constraint shown in Eq.~(\ref{throughput_memory})  and privacy constraint in Eq.~(\ref{throughput_init}).  

\begin{equation}\label{throughput_memory}
	Req_{i \rightarrow m} \leq Mem_{j} 
\end{equation} 
\vspace{-1em}
\begin{equation}\label{throughput_init}
	g(1, 1, 0) = t_{comp}^{0,0}
\end{equation} 

Algo.~\ref{strategy} describes the pseudo-code to find the optimal solution $T_{throu}^{opt}$ and the corresponding model partition and allocation strategy. In Algo.~\ref{strategy}, we first initialize the dynamic programming table $g(m, S', j)$ and choice table $choice(m, S, j)$, and assign $t_{comp}^{0,0}$ to $g(1, 1, 0)$ (lines 1-2). We then traverse the layers of the large language model from layer $1$. For each layer, we traverse all the computing devices with sufficient memory and calculate the maximum latency (lines 3-23). After filling the DP table, we can find the maximum latency, based on which we then backtrace the choice table and finally get the model partition and allocation strategy (lines 24-32). The computational complexity of Algo.~\ref{strategy} is $O(N^{2} \times 2^{M} \times M^{2})$, where $N$ is the number of layers of the LLM model and $M$ is the number devices in the network.

\begin{algorithm}[t]
	\caption{Joint device selection and LLM partition for optimizing throughput}\label{strategy}
	\KwIn{A LLM model; Computing devices $M$; Profiled traces; bandwidth $B_{k,j}$;}
	\KwOut{the device selection and LLM partition strategy $R$ }
	\tcp{initialization}
	Initialize DP table $g(i, S, k)=INF$, and choice table $choice(m, S, j)=NULL$ to record the strategy\;
	Enforce first layer to be allocated to node $0$ by $g(1, 1, 0) = t_{comp}^{0,0}$ and $choice(1, 1, 0)=(0,0,0)$\;
	\tcp{fill in DP table}
	\For{$i=1$ to $N-1$}{
		\For{each subset $S \subseteq M$}{
			\For{last node $k \in S$}{
				\For{$m=i+1$ to $N-1$}{
					\For{$j \in M \setminus S$}{
						\If{$Mem_{j} \leq \sum_{i}^{m} Req_{i}$}{
							Continue\;
						}
						\Else{
							Get $S'$ by adding node $j$ to the selected device set $S$\;
							Calculate current mamixum execution time $T_{max}$ via Eq.~(\ref{transistion_equation}) for the maximum execution time in all stages\;
						}
						\If{$T_{max} \leq g(i,S,k)$}{
							$g(m,S',j) = T_{max}$\;
							Record the current strategy $choice(m,S',j) = (i,j,k)$\;
						}
					}
				}
			}

		}
		
	}
	\tcp{backtrace for optimal allocation}
	Initialize optimal strategy $R$\;
	Find selected device set $S$ and the last selected node $N_{last}$ by $S, N_{last} = argmin_{S,k} (g(N-1, S, k))$\;
	Initialize $layer = N-1$\;
	\While{$layer > 0$}{
		$(i, j, k)=choice(layer, S, N_{last})$\;
		Add $(i \rightarrow layer, j)$ to $R$\;
		Update $layer, S$ and $N_{last}$\;
	}

	\Return $R$\;
\end{algorithm}

\textbf{Pipeline Execution Optimization.}
Note that the above problem formulation and solution are based on the ideal case, where there is no idle device at any time. A device processes a batch of data and continues to handle another batch of data without waiting. However, it is impractical for LLM inference in real-world cases.

\begin{figure*}[t]
	\centering
	\subfigure [Bubbles]{
		\includegraphics[width=0.8\linewidth]{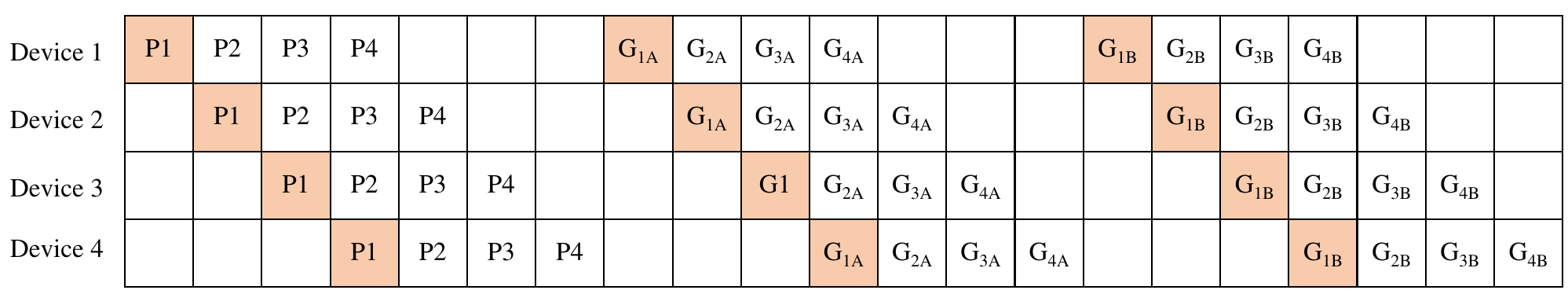}
	}
	
	\subfigure [No-bubbles]{
		\includegraphics[width=0.8\linewidth]{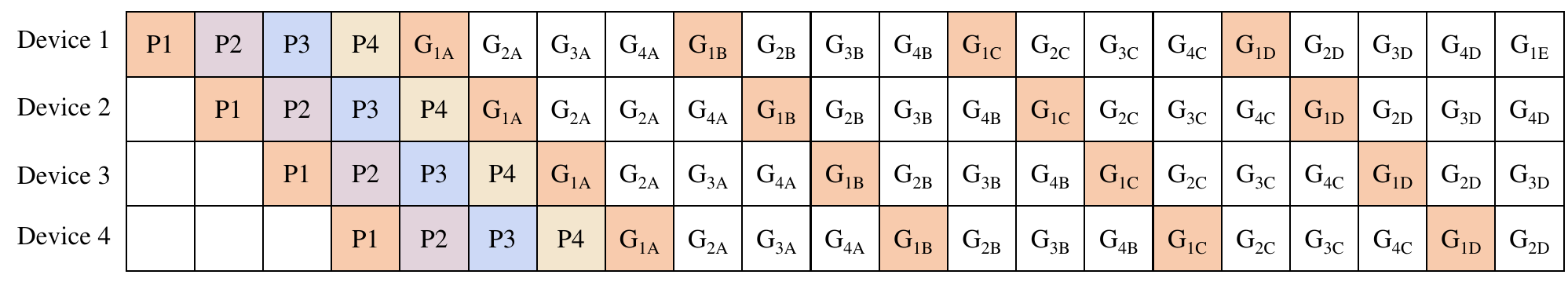}
	}
	\caption{Different pipeline execution strategies of EdgeShard. EdgeShard-No-bubbles reduces device idle time to improve throughput by allowing immediate token generation of a micro-batch without waiting for other micro-batches.}
	\label{pipeline-execution}
	%\vspace{-0.3cm}
\end{figure*}

As shown in Fig.~\ref{pipeline-execution}(a), different from those one-phase computation applications, the decoder-based LLM application has an autoregressive nature, where there will be multiple tokens to be generated and the calculation of the current token relies on all the previous tokens. The computation of the current token cannot start until it gets the previously generated token. It leads to bubbles in pipeline execution.

To approximate the ideal case and enhance the resource utilization for improving throughput, we tend to reduce the bubbles in the pipeline execution. We propose EdgeShard-No-bubbles, which allows for immediate token generation without waiting for the ending of all micro-batches in an iteration. As shown in Fig.~\ref{pipeline-execution}(b), after the prefill stage $P1$ ends of the first batch, Device 1 immediately executes the token generation of the first batch as indicated by $G_{1A}$. Similarly, when $G_{1A}$ ends, Device 1 goes to the next iteration of token generation indicated by $G_{1B}$. Compared to EdgeShard-Bubbles, EdgeShard-No-bubbles reduces bubbles by mitigating device idle time and is expected to improve throughput. From the pipeline execution graph in Fig.~\ref{pipeline-execution}, we can see that EdgeShard-No-bubbles generates more tokens at the same time.

\section{Experimental Evaluation}

\subsection{Experimental Setup}
\begin{figure}[t]
	\centering
	\includegraphics[width=0.8\linewidth]{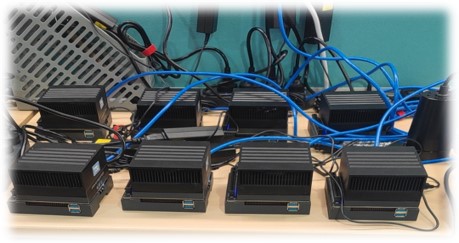}
	\caption{Our testbed has heterogeneous edge devices and cloud server. Their specifications are shown in Table \ref{t:physical-devices}.}
	\label{f:testbed}
\end{figure}

\textbf{Testbed.} We use various edge devices and cloud servers to act as the heterogeneous computation devices in collaborative edge computing. The specifications of those devices are listed in Table.~\ref{t:physical-devices}. We use 15 devices, including 12 Jetson AGX Orin, 2 Jetson Orin NX, and one cloud server to configure the collaborative edge network. The physical testbed is shown in Fig.~\ref{f:testbed}. Those devices are connected with a route and a switch. The bandwidth between any two devices is 1000Mbps. We use the Linux TC tool \cite{hubert2002linux} to vary network bandwidth and communication latency between devices.

\begin{table}[t]
	\centering
	\caption{Specifications of heterogeneous physical devices}
	\label{t:physical-devices} 
	\begin{tabular}{cccc}
		\toprule
		Category & Device & Memory & AI Performance \\
		\toprule
		 Edge Device & Jetson AGX Orin & $32$GB & $3.33$ TFLOPS  \\
		 Edge Device & Jetson Orin NX & $16$GB & $1.88$ TFLOPS \\
		 Cloud Server & RTX 3090 & $24$GB & $36$ TFLOPS \\
		\bottomrule
	\end{tabular}
\end{table}

\textbf{Benchmarks.} We test the performance of EdgeShard with a series of Llama2 models \cite{touvron2023Llama}, including Llama2-7B, Llama2-13B, and Llama2-70B. Llama2 is released by Meta in July 2023 and is one of the most popular and powerful open-source large language models, representing a groundbreaking leap in the field of artificial intelligence and natural language processing. For the model inference, we adopt the text generation task to test the performance. We use the WikiText-2 dataset \cite{merity2017pointer} from HuggingFace. We extract a subset of samples with the length of input tokens as 32 and generate 96 tokens. We use full-precision model inference in all the following experiments.

\textbf{Baselines.} We compare the performance in terms of latency and throughput of EdgeShard with various baselines. (We don't use the cloud-only as a baseline because it requires the input token to be transmitted to the cloud server, which may lead to privacy concerns).
%the GPUs of cloud servers do not have enough memory to accommodate the inference tasks, even for the smallest Llama2-7B model).
\begin{itemize}
	\item \textbf{Edge-Solo.} In this case, the LLMs are deployed locally on an edge device without model partition.
	\item \textbf{Cloud-Edge-Even.} In this case, the LLMs are evenly partitioned into two parts. One is allocated to the edge device, and another is allocated to the cloud server.
	\item \textbf{Cloud-Edge-Opt.} In this case, the LLMs are partitioned into two shards. One is allocated to the edge device, and another is allocated to the cloud server. For the partition strategy of LLMs, we also use the proposed dynamic programming algorithms. The difference is that there is only two devices as the algorithm input.
\end{itemize}

\begin{table*}[t]
	\caption{Performance of LLM Inference. (Average Latency: milliseconds/token; Throughput: tokens/second).}
	\centering
	\label{t:overall_performance} 
	\begin{tabular}{lcccccc}
		\hline
		& \multicolumn{2}{c}{Llama2-7B} & \multicolumn{2}{c}{Llama2-13B} & \multicolumn{2}{c}{Llama2-70B}  \\ \cmidrule(r){2-3} \cmidrule(r){4-5} \cmidrule(r){6-7}  
		&     Latency      &    Throughput     &    Latency      &     Throughput     &     Latency      &    Throughput       \\ \hline
		Edge-Solo &     140.34      &    24.36      &    OOM       &    OOM     &     OOM      &    OOM       \\
		Cloud-Edge-Even &    227.35       &    7.56      &     319.44      &     4.68    &     OOM      &    OOM    \\ 
		Cloud-Edge-Opt &     140.34      &    24.36      &     243.45      &     4.74     &    OOM       &    OOM \\
		EdgeShard &    75.88       &    52.45      &     173.43      &     10.45    &     3086.43      &    1.25    \\ \hline
	\end{tabular}
\end{table*}

\begin{figure*}[t]
	\centering
	\subfigure[Llama2-7B]{
		\begin{minipage}[t]{0.3\linewidth}
			\centering
			\includegraphics[width=\linewidth]{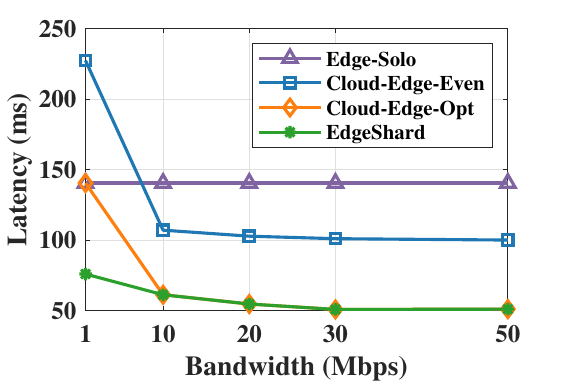}
		\end{minipage}
	}\hspace{-3mm}
	\subfigure[Llama2-13B]{
		\begin{minipage}[t]{0.3\linewidth}
			\centering
			\includegraphics[width=\linewidth]{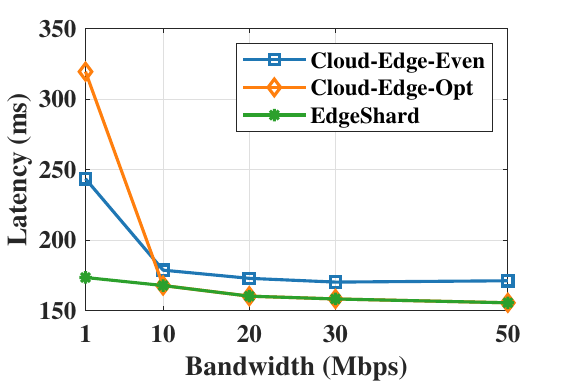}
		\end{minipage}
	}\hspace{-3mm}
	\subfigure[Llama2-70B]{
		\begin{minipage}[t]{0.3\linewidth}
			\centering
			\includegraphics[width=\linewidth]{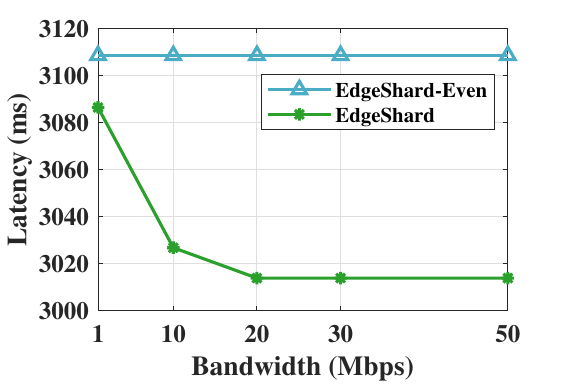}
		\end{minipage}
	}
	\centering
	\caption{Impact of Network Bandwidth to Latency of Collaborative LLMs inference}
	\label{exp: bandwidth_latency}
	\vspace{-0.3cm}
\end{figure*}

\begin{figure*}[t]
	\centering
	\subfigure[Llama2-7B]{
		\begin{minipage}[t]{0.3\linewidth}
			\centering
			\includegraphics[width=\linewidth]{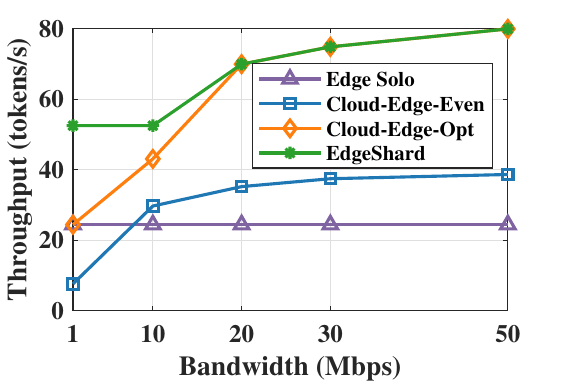}
		\end{minipage}
	}\hspace{-3mm}
	\subfigure[Llama2-13B]{
		\begin{minipage}[t]{0.3\linewidth}
			\centering
			\includegraphics[width=\linewidth]{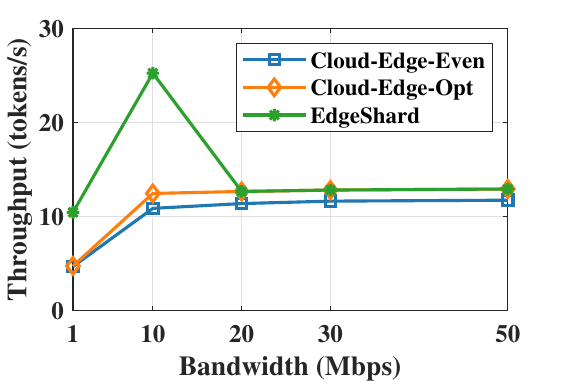}
		\end{minipage}
	}\hspace{-3mm}
	\subfigure[Llama2-70B]{
		\begin{minipage}[t]{0.3\linewidth}
			\centering
			\includegraphics[width=\linewidth]{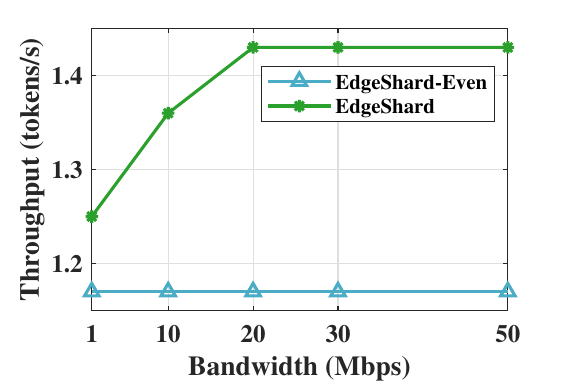}
		\end{minipage}
	}
	\centering
	\caption{Impact of Bandwidth to Throughput of Collaborative LLMs Inference}
	\label{exp: bandwidth_throughput}
	\vspace{-0.3cm}
\end{figure*}

\subsection{Overall Evaluation}
We set AGX Orin as the source node and the bandwidth between the source node and the cloud server as $1$Mbps. The bandwidth between other computing devices is set to be 50Mbps with a variance of 20\%. To test the throughput, we set the batch size as the maximum batch size that the participating devices can support. The latency and throughput of LLM inference are shown in Table.~\ref{t:overall_performance}. 

We have the following observations. First, EdgeShard is potential and beneficial for large language model deployment. For Llama2-70B model, the memory requirement is about 280GB, which far exceeds the memory capacity of solo edge deployment and cloud-edge collaborative deployment. They will have the out-of-memory issue (OOM). However, EdgeShard tackles this challenge by splitting the large model into shards and allocating them to multiple devices, enabling collaborative model inference. 
Second, EdgeShard achieves obviously lower inference latency and higher inference throughput than baseline methods. For Llama2-7B model, EdgeShard achieves 75.88ms latency, which is about 1.85x faster than Edge-Solo and Cloud-Edge-Opt, and about 3x faster than Cloud-Edge-Even. For the inference throughput, EdgeShard achieves 52.45 tokens per second with a maximum batch size of 8, which is around 2.2 times larger than Edge-Solo and Cloud-Edge-Opt, and about 7 times larger than Cloud-Edge-Even. Similar performance improvement is also observed for Llama2-13B model, where EdgeShard achieves 45.7\% and 28.8\% lower latency than Cloud-Edge-Even and Cloud-Edge-Opt, respectively. Also, EdgeShard has 2.23x and 2.2x higher throughput than Cloud-Edge-Even and Cloud-Edge-Opt. 
Third, we can also see that, for Llama2-7B, Cloud-Edge-Opt tends to have the same performance in terms of both inference latency and throughput as Edge-Solo. This is because the bandwidth between the source node and the cloud server is very limited in this experimental setting, i.e., 1Mbps. The optimal deployment strategy of Cloud-Edge-Collaboration is local execution, which is the same as Edge-Solo.  

\subsection{Effects of Bandwidth}
We set the source node as AGX Orin and vary the bandwidth between the cloud server and the source node from 1Mbps to 50Mbps. The performance of the latency and throughput of LLM inference are shown in Fig.~\ref{exp: bandwidth_latency} and Fig.~\ref{exp: bandwidth_throughput}, respectively. 

For Llama2-13B, a single AGX Orin cannot accommodate the full model. We only compare the performance among Cloud-Edge-Even, Cloud-Edge-Opt, and EdgeShard. Similarly, due to the memory constraint, the three baseline methods are not able to deploy the Llama2-70B model. Instead, we compare the performance of EdgeShard with its variant, i.e., EdgeShard-Even, where the model is equally partitioned and deployed to all the participating computing devices. It selects 11 AGX Orin and 1 RTX 3090 to deploy the Llama2-70B model.

In terms of latency, except for Edge-Solo, the latency of the other three methods decreases with the increasing bandwidth. This is because the three methods are collaboration-based, and the latency is influenced by the data transmission time. The increasing bandwidth leads to reduced communication time. We can also see that for the collaboration methods, there is a dramatically latency reduction when the cloud-source bandwidth changes from 1Mbps to 10Mbps and a minor variance from 10Mbps to 50Mbps. This is because the bandwidth is gradually saturated at that time, and the computation time becomes the bottleneck. 

Moreover, we can see that when the bandwidth is greater than 10Mbps, cloud-edge collaboration methods outperform the Edge-Solo method, as the cloud-edge collaboration methods introduce the powerful cloud server for computation acceleration. However, when the bandwidth is 1Mbps, Cloud-Edge-Even performs worse than EdgeSolo. This is because the data transmission cost is high in this case. The Cloud-Edge-Opt method tends to deploy the LLM model locally, which is the same as the Edge-Solo method. Interestingly, the latency of Cloud-Edge-Opt and EdgeShard is nearly the same when the bandwidth is greater than 10Mbps. We found that EdgeShard generates the same model partition and allocation policies as the Cloud-Edge-Opt method. The variance comes from the small fluctuations in model execution. It shows that the performance of EdgeShard will not be worse than that of Cloud-Edge-Opt, and the Cloud-Edge-Opt method is a special case of EdgeShard. A similar pattern is also observed for Llama2-13B. For Llama2-70B, EdgeShard performs better than its variant EdgeShard-Even, as there is resource heterogeneity among cloud server and edge devices, and EdgeShard adaptively partitions the LLMs among computing devices. However, the performance improvement is not so obvious as there are 11 AGX with the same computation capacity and only 1 RTX 3090.

In terms of throughput, similar patterns to the latency evaluation are also found for Llama2-7B model. Differently and interestingly, for Llama2-13B, EdgeShard does not show a closing performance with the Cloud-Edge-Opt method when the bandwidth is 10Mbps, but with a great improvement, where EdgeShard has about 2x higher throughput than the Cloud-Edge-Opt method. This is because of the high memory consumption of the RTX 3090 and the source node, i.e., AGX Orin. We observed that for the Cloud-Edge-Opt, the memory consumption of the two devices goes up to 95\% and 98\%, respectively, which only allows for a maximum batch size of 4. Otherwise, there will not be enough memory for the KV cache on the computing devices. However, when the bandwidth is 10Mbps, EdgeShard involves several edge devices where the memory consumption of an individual device becomes dramatically decreased, allowing for a larger batch size, i.e., 8 in this case. When the bandwidth is higher than 10Mbps, EdgeShards tends to have the same model partition and allocation strategy as Cloud-Edge-Opt, which yields a closing performance, as shown in Llama2-7B. For Llama2-70B, there is a slight throughput improvement of EdgeShard, and EdgeShard-Even shows a steady throughput as the evenly partition strategy will not change with the cloud-source bandwidth.

\subsection{Effects of Source Node}
\begin{figure}[t]
 \centering
 \subfigure[Llama2-7B - Latency]{
	 \includegraphics[width=0.9\linewidth]{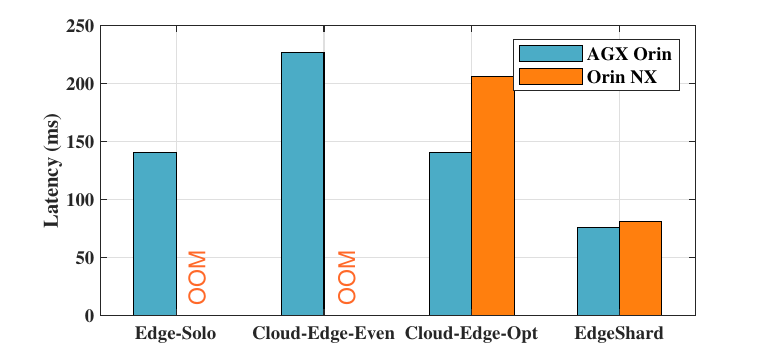}
	 %\caption{fig1}
	 }
 \subfigure[Llama2-7B - Throughput]{
	 \includegraphics[width=0.9\linewidth]{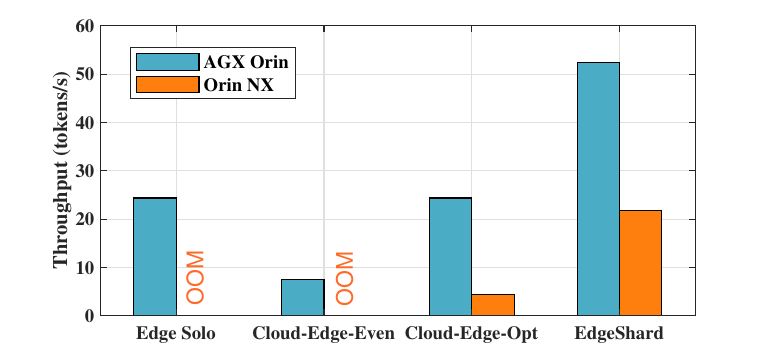}
	 }
 \caption{Impact of Source Node}
 \label{exp-source}
\end{figure}

We also test the influence of the source node on the inference latency and throughput, as the source node may have different computation and memory capacities, and EdgeShard enforces the first layer of LLM models residing on the source node to avoid raw data transmission. We set the source node as AGX Orin and Orin NX, respectively, and compare their performance. We set the bandwidth between the source node and the cloud server as 1Mbps. The results of Llama2-7B inference are shown in Fig.~\ref{exp-source}.

We find that when the source node is Orin NX, the Edge-Solo and Cloud-Edge-Even methods encounter the OOM error. This is due to the relatively lower memory of Orin NX, which cannot accommodate the Llama2-7B model, even for half part of the model. The difference between the two cases under the Cloud-Edge-Opt method is much more obvious than that of EdgeShard. For Cloud-Edge-Opt, there is about a 60ms gap, and for EdgeShard, the gap is about 5ms. This is because there are only two devices in the Cloud-Edge-Opt case, and it tends to put more layers on the source node. However, AGX Orin is much more powerful than Orin NX in terms of computation capacity. EdgeShard tends to involve more devices and put fewer model layers on the source node, which can fill in the gap in computation capacity between the source nodes. A similar phenomenon is also observed for the throughput, where AGX Orin has 6x higher throughput than Orin Nx for the Cloud-Edge-Opt method and only 2x higher throughput under the EdgeShard method. It shows EdgeShard can make full use of the computation resources in the network to optimize the performance.   

\subsection{Effects of Pipeline Execution strategy}
\begin{figure}[t]
	\centering
	\subfigure[Llama2-7B - Throughput]{
		\includegraphics[width=0.83\linewidth]{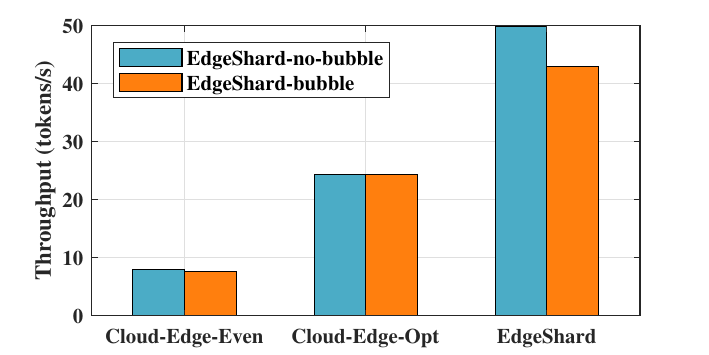}
		%\caption{fig1}
	}
	\subfigure[Llama2-13B - Throughput]{
		\includegraphics[width=0.83\linewidth]{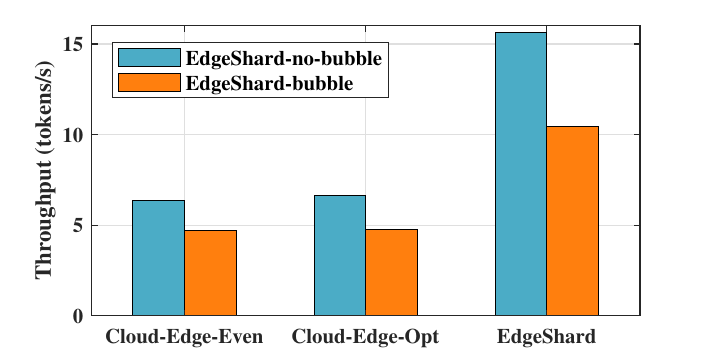}
	}
	\caption{Impact of Pipeline Execution Strategy}
	\label{exp: pipeline-strategy}
\end{figure}

We evaluate the two pipeline execution strategies. We set the bandwidth between the cloud server and the source node as 1Mbps. The results are shown in Fig.~\ref{exp: pipeline-strategy}.

We can see that for all methods, EdgeShard-No-bubble outperforms EdgeShard-Bubble. Specifically, for Llama2-7b, EdgeShard-No-bubble achieves an improvement of about 0.36 and 6.96 tokens per second than Edgeshard-bubble for Cloud-Edge-Even and EdgeShard, respectively. For the Cloud-Edge-Opt method, it selects local execution in this case. There is no pipeline execution, so the throughput for the two methods is the same. For Llama2-13b, EdgeShard-No-bubble achieves an improvement of about 1.69, 1.89, and 5.21 tokens per second than Edgeshard-Bubble for Cloud-Edge-Even, Cloud-Edge, and EdgeShard, respectively. Compared to EdgeShard-Bubble, EdgeShard-No-bubble does not need to wait for the completion of all micro-batches in an iteration and can effectively reduce the devices' idle time, thus leading to a higher throughput.

\section{Related Work}\label{sec: literature}
This section reviews research works of LLM in the edge computing environment from two aspects, i.e., edge computing for efficient LLM deployment and LLM for optimizing edge computing.

\subsection{Edge Computing for Efficient LLM}
LLM is computation-intensive and memory-consuming. To address the issue of memory wall, quantization is widely adopted \cite{shen2023agile,frantar2022gptq,frantar2022optq,lin2023awq,xiao2023smoothquant,shen2024edgeqat}. GPTQ \cite{frantar2022gptq} quantizes LLM with hundreds of billions of parameters to 3-4bits based on approximate second-order information.
Lin et al. \cite{lin2023awq} reduce quantization error by optimizing channel scaling to preserve the salient important weights. They are weight-only quantization. SmoothQuant \cite{xiao2023smoothquant} and Agile-Quant \cite{shen2023agile} take a further step, which quantize not only the model weights, but also the model activations. However, the computation capacity and memory of a single device is still limited even for quantized LLM. Moreover, the performance of quantized LLM usually cannot be compared to that of its full-size model. 

Other works \cite{wang2023privatelora,chen2023netgpt} tend to leverage the cloud-edge collaboration to partition and distribute the massive computation workload of LLM inference and finetuning. Wang et al. \cite{wang2023privatelora} increase the throughput by distributing the computation between cloud servers and edge devices, and reducing the communication overhead of transmitting the activations between the central cloud and edge devices by leveraging the low-rank property of residual activations. Chen et al. \cite{chen2023netgpt} efficiently leverage location-based information of edge devices for personalized prompt completion during collaborative edge-cloud LLM serving. However, the latency between edge devices and the central cloud is usually high and unstable, which will affect the inference and finetuning performance of LLM. 

Our work is different from those works. We propose a general framework to integrate the computation resources of heterogeneous and ubiquitous cloud servers and edge devices. The framework allows the adaptive selection of computation devices and partition of the computation workload of LLM inference for optimized latency and throughput.
 
\subsection{LLM for Optimizing Edge Computing}
LLMs also have great potential in making complex and coherent decisions. There are also some works that leverage LLM to optimize resource utilization in edge computing, such as resource allocation and task offloading, network management, and intelligent IoT control. Li et al. \cite{dong2023lambo} propose LAMBO, a LLM-based task offloading framework for mobile edge computing, to address the challenging issues of heterogeneous constraints, partial status perception, diverse optimization objectives, and dynamic environment that are not well addressed in traditional task offloading research. LAMBO shows that LLM is more effective compared to traditional DNN and deep reinforcement learning-based methods in complex and dynamic edge computing environments. They further design a LLM-based multi-agent system and incorporate communication knowledge and tools into the system, empowering it with the ability to optimize semantic communication in a 6G network \cite{jiang2023large}. Apart from optimization of resource utilization, Shen et al. \cite{shen2024large} leverage the outstanding abilities of GPT in language understanding and code generation to train new models among federated edge devices. Rong et al. \cite{rong2024leveraging} leverage LLMs to generate adaptive control algorithms for addressing the diverse, dynamic, and decentralized network conditions in 6G integrated terrestrial network (TN) and non-terrestrial network (NTN). Though LLMs have shown great potential in making intelligent decisions, especially in complex and dynamic edge computing systems, the related research is still in the early stages. Challenges such as significant resource consumption, latency of decision-making, and uncertainty of generated decisions need further studies.

\section{Discussion and Future Works}
This section discusses some open issues and future works that may appeal to readers.

%Relationship with Computing Power Network.

\textbf{Incentive mechanisms.} In this work, we partition the LLM into multiple shards and allocate them to heterogeneous devices. For edge computing scenarios, such as smart home and smart factory, there is a set of trusted devices owned by a single stakeholder. They may be able to use those devices for collaborative inference. However, if the devices belong to different stakeholders, they may not be willing to share devices' computation resources. Further incentive mechanisms are needed to reward resource sharing.

\textbf{Batch size aware optimization.} Large batch size will increase memory usage and affect the inference throughput. As shown in the experiment, by partitioning the workload of LLM inference to multiple devices, the memory usage of participating devices can be reduced and thus allows for a larger batch size, leading to increased throughput. However, the designed dynamic programming algorithm does not consider the influence of batch size, which remains space for further optimization.

\section{Conclusion}\label{sec: conclusion}
In this work, we propose EdgeShard to enable the efficient deployment and distributed inference of LLMs on collaborative edge devices and cloud servers. We formulate a joint device selection and model partition problem to optimize inference latency and throughput, respectively, and solve it using dynamic programming algorithms. Experimental results show that edgesplit can adaptively determine the LLM partition and deployment strategy under various heterogeneous network conditions for optimizing inference performance. Edgeshard is not designed to replace cloud-based LLM inference, but to provide a flexible and adaptive LLM serving methods by utilizing ubiquitous computing devices. Experiments also shows that EdgeShard outperforms the cloud-edge collaborative inference method when cloud bandwidth is insufficient and tends to yield the same deployment strategy as the cloud-edge collaborative inference method when facing relatively abundant cloud bandwidth. 

This is a pioneering work of deploying LLM in collaborative edge computing environment. We hope this work can stimulate more ideas and further research in this promising area.

\section{Acknowledgement}
This work was supported by the Research Institute for Artificial Intelligence of Things, The Hong Kong Polytechnic University, HK RGC Grant for Theme-based Research Scheme No. T43-513/23-N, and National Natural Science Foundation of China and Hong Kong RGC Collaborative Research Scheme No. CRS\_PolyU501-23.

%\newpage

\bibliographystyle{IEEEtran}
\bibliography{main.bib}

\begin{thebibliography}{10}
\providecommand{\url}[1]{#1}
\csname url@samestyle\endcsname
\providecommand{\newblock}{\relax}
\providecommand{\bibinfo}[2]{#2}
\providecommand{\BIBentrySTDinterwordspacing}{\spaceskip=0pt\relax}
\providecommand{\BIBentryALTinterwordstretchfactor}{4}
\providecommand{\BIBentryALTinterwordspacing}{\spaceskip=\fontdimen2\font plus
\BIBentryALTinterwordstretchfactor\fontdimen3\font minus \fontdimen4\font\relax}
\providecommand{\BIBforeignlanguage}[2]{{%
\expandafter\ifx\csname l@#1\endcsname\relax
\typeout{** WARNING: IEEEtran.bst: No hyphenation pattern has been}%
\typeout{** loaded for the language `#1'. Using the pattern for}%
\typeout{** the default language instead.}%
\else
\language=\csname l@#1\endcsname
\fi
#2}}
\providecommand{\BIBdecl}{\relax}
\BIBdecl

\bibitem{achiam2023gpt}
J.~Achiam, S.~Adler, S.~Agarwal, L.~Ahmad, I.~Akkaya, F.~L. Aleman, D.~Almeida, J.~Altenschmidt, S.~Altman, S.~Anadkat \emph{et~al.}, ``Gpt-4 technical report,'' \emph{arXiv preprint arXiv:2303.08774}, 2023.

\bibitem{touvron2023Llama}
H.~Touvron, L.~Martin, K.~Stone, P.~Albert, A.~Almahairi, Y.~Babaei, N.~Bashlykov, S.~Batra, P.~Bhargava, S.~Bhosale \emph{et~al.}, ``Llama 2: Open foundation and fine-tuned chat models,'' \emph{arXiv preprint arXiv:2307.09288}, 2023.

\bibitem{anil2023palm}
R.~Anil, A.~M. Dai, O.~Firat, M.~Johnson, D.~Lepikhin, A.~Passos, S.~Shakeri, E.~Taropa, P.~Bailey, Z.~Chen \emph{et~al.}, ``Palm 2 technical report,'' \emph{arXiv preprint arXiv:2305.10403}, 2023.

\bibitem{vaswani2017attention}
A.~Vaswani, N.~Shazeer, N.~Parmar, J.~Uszkoreit, L.~Jones, A.~N. Gomez, {\L}.~Kaiser, and I.~Polosukhin, ``Attention is all you need,'' \emph{Advances in neural information processing systems}, vol.~30, 2017.

\bibitem{shi2016edge}
W.~Shi, J.~Cao, Q.~Zhang, Y.~Li, and L.~Xu, ``Edge computing: Vision and challenges,'' \emph{IEEE Internet of Things Journal}, vol.~3, no.~5, pp. 637--646, 2016.

\bibitem{chen2019deep}
J.~Chen and X.~Ran, ``Deep learning with edge computing: A review,'' \emph{Proceedings of the IEEE}, vol. 107, no.~8, pp. 1655--1674, 2019.

\bibitem{shen2023agile}
X.~Shen, P.~Dong, L.~Lu, Z.~Kong, Z.~Li, M.~Lin, C.~Wu, and Y.~Wang, ``Agile-quant: Activation-guided quantization for faster inference of llms on the edge,'' \emph{arXiv preprint arXiv:2312.05693}, 2023.

\bibitem{frantar2022gptq}
E.~Frantar, S.~Ashkboos, T.~Hoefler, and D.~Alistarh, ``Gptq: Accurate post-training quantization for generative pre-trained transformers,'' \emph{arXiv preprint arXiv:2210.17323}, 2022.

\bibitem{frantar2022optq}
------, ``Optq: Accurate quantization for generative pre-trained transformers,'' in \emph{The Eleventh International Conference on Learning Representations}, 2022.

\bibitem{lin2023awq}
J.~Lin, J.~Tang, H.~Tang, S.~Yang, X.~Dang, and S.~Han, ``Awq: Activation-aware weight quantization for llm compression and acceleration,'' \emph{arXiv preprint arXiv:2306.00978}, 2023.

\bibitem{xiao2023smoothquant}
G.~Xiao, J.~Lin, M.~Seznec, H.~Wu, J.~Demouth, and S.~Han, ``Smoothquant: Accurate and efficient post-training quantization for large language models,'' in \emph{International Conference on Machine Learning}.\hskip 1em plus 0.5em minus 0.4em\relax PMLR, 2023, pp. 38\,087--38\,099.

\bibitem{shen2024edgeqat}
X.~Shen, Z.~Kong, C.~Yang, Z.~Han, L.~Lu, P.~Dong, C.~Lyu, C.-h. Li, X.~Guo, Z.~Shu \emph{et~al.}, ``Edgeqat: Entropy and distribution guided quantization-aware training for the acceleration of lightweight llms on the edge,'' \emph{arXiv preprint arXiv:2402.10787}, 2024.

\bibitem{wang2023privatelora}
Y.~Wang, Y.~Lin, X.~Zeng, and G.~Zhang, ``Privatelora for efficient privacy preserving llm,'' \emph{arXiv preprint arXiv:2311.14030}, 2023.

\bibitem{chen2023netgpt}
Y.~Chen, R.~Li, Z.~Zhao, C.~Peng, J.~Wu, E.~Hossain, and H.~Zhang, ``Netgpt: A native-ai network architecture beyond provisioning personalized generative services,'' 2023.

\bibitem{zhang2022eaas}
M.~Zhang, J.~Cao, Y.~Sahni, Q.~Chen, S.~Jiang, and T.~Wu, ``Eaas: A service-oriented edge computing framework towards distributed intelligence,'' in \emph{2022 IEEE International Conference on Service-Oriented System Engineering (SOSE)}.\hskip 1em plus 0.5em minus 0.4em\relax IEEE, 2022, pp. 165--175.

\bibitem{zhang2022ents}
M.~Zhang, J.~Cao, L.~Yang, L.~Zhang, Y.~Sahni, and S.~Jiang, ``Ents: An edge-native task scheduling system for collaborative edge computing,'' in \emph{2022 IEEE/ACM 7th Symposium on Edge Computing (SEC)}.\hskip 1em plus 0.5em minus 0.4em\relax IEEE, 2022, pp. 149--161.

\bibitem{huang2019gpipe}
Y.~Huang, Y.~Cheng, A.~Bapna, O.~Firat, D.~Chen, M.~Chen, H.~Lee, J.~Ngiam, Q.~V. Le, Y.~Wu \emph{et~al.}, ``Gpipe: Efficient training of giant neural networks using pipeline parallelism,'' \emph{Advances in neural information processing systems}, vol.~32, 2019.

\bibitem{narayanan2019pipedream}
D.~Narayanan, A.~Harlap, A.~Phanishayee, V.~Seshadri, N.~R. Devanur, G.~R. Ganger, P.~B. Gibbons, and M.~Zaharia, ``Pipedream: Generalized pipeline parallelism for dnn training,'' in \emph{Proceedings of the 27th ACM Symposium on Operating Systems Principles}, 2019, pp. 1--15.

\bibitem{devlin2018bert}
J.~Devlin, M.-W. Chang, K.~Lee, and K.~Toutanova, ``Bert: Pre-training of deep bidirectional transformers for language understanding,'' \emph{arXiv preprint arXiv:1810.04805}, 2018.

\bibitem{hubert2002linux}
B.~Hubert \emph{et~al.}, ``Linux advanced routing \& traffic control howto,'' \emph{Netherlabs BV}, vol.~1, pp. 99--107, 2002.

\bibitem{merity2017pointer}
S.~Merity, C.~Xiong, J.~Bradbury, and R.~Socher, ``Pointer sentinel mixture models,'' in \emph{International Conference on Learning Representations}, 2017.

\bibitem{dong2023lambo}
L.~Dong, F.~Jiang, Y.~Peng, K.~Wang, K.~Yang, C.~Pan, and R.~Schober, ``Lambo: Large language model empowered edge intelligence,'' \emph{arXiv preprint arXiv:2308.15078}, 2023.

\bibitem{jiang2023large}
F.~Jiang, L.~Dong, Y.~Peng, K.~Wang, K.~Yang, C.~Pan, D.~Niyato, and O.~A. Dobre, ``Large language model enhanced multi-agent systems for 6g communications,'' \emph{arXiv preprint arXiv:2312.07850}, 2023.

\bibitem{shen2024large}
Y.~Shen, J.~Shao, X.~Zhang, Z.~Lin, H.~Pan, D.~Li, J.~Zhang, and K.~B. Letaief, ``Large language models empowered autonomous edge ai for connected intelligence,'' \emph{IEEE Communications Magazine}, 2024.

\bibitem{rong2024leveraging}
B.~Rong and H.~Rutagemwa, ``Leveraging large language models for intelligent control of 6g integrated tn-ntn with iot service,'' \emph{IEEE Network}, 2024.

\end{thebibliography}
\end{document}